\documentclass[%
aps,prl,reprint,showpacs,
 amsmath,amssymb,nofootinbib,nobibnotes,superscriptaddress
]{revtex4-1}
\usepackage{blindtext}
 \usepackage{ulem}
    \usepackage{float}
\usepackage{graphicx}
\usepackage{dcolumn}
\usepackage[font=small,labelfont=bf,
   justification=justified,
   format=plain]{caption} 

\usepackage{lipsum}

\usepackage[colorlinks=true, linkcolor = red, citecolor = blue]{hyperref}
\usepackage{subcaption}

\usepackage{float}
\usepackage{xcolor}

\usepackage{ulem}
\begin{document}
\newcommand{\lu}[1]{\textcolor{red}{#1}}
\newcommand{\quita}[1]{\textcolor{orange}{#1}}
\newcommand{\JCH}[1]{\textcolor{blue}{#1}}
\title{New mechanism for primordial black hole formation during reheating}

\author{Luis E. Padilla}
\email{lepadilla@icf.unam.mx}
\affiliation{Instituto de Ciencias Físicas, Universidad Nacional Autónoma de México,
62210, Cuernavaca, Morelos, México.}
\affiliation{Mesoamerican Centre for Theoretical Physics,
Universidad Aut\'onoma de Chiapas, Carretera Zapata Km. 4, Real
del Bosque (Ter\'an), Tuxtla Guti\'errez 29040, Chiapas, M\'exico.}

\author{Juan Carlos Hidalgo}
  \email{hidalgo@icf.unam.mx}
\affiliation{Instituto de Ciencias Físicas, Universidad Nacional Autónoma de México,
62210, Cuernavaca, Morelos, México.}

\author{Karim A. Malik}
\email{k.malik@qmul.ac.uk}
\affiliation{Astronomy Unit, Queen Mary University of London, London, E1 4NS, United Kingdom.}

\date{\today}%

\begin{abstract}
In the scalar field dark matter model virialized halos present condensed central cores called boson stars. Considering the equivalent process during reheating, we look at the formation of primordial black holes (PBHs) through the gravitational collapse of structures virialized in this era. We present the criteria necessary for collapse of either the whole structure, or that of the central core, in terms of the threshold amplitude for the primordial density contrast. This is computed for both the free and the self-interacting scalar fields. We discuss the relevance of our results for the abundance of PBHs.
 \end{abstract}

\maketitle

\section{Introduction}
For quite some time scalar fields have been considered key constituents of the
universe and the Einstein-Klein-Gordon system of equations is
therefore commonly used to describe the dynamics of a universe
governed by a single or multiple components of such fields. Scalar
fields play crucial roles in the dark sector: they are strong
candidates for the Dark Matter -- scalar field dark matter (SFDM)
\citep{Matos:2000ng,hui2017ultralight} -- while quintessence models
can account for Dark Energy
(see e.g.~Ref.~\cite{quintessence}). Finally, and arguably most importantly, scalar fields are regarded as the natural
realization of the primordial inflationary mechanism
\citep{Guth,Linde1982}.  Inflation requires a period of transition to
the standard hot Big Bang, a process called reheating (for reviews see Refs.~\cite{reh2,Lozanov:2019jxc}).  Energy
transfer during reheating is most effective when scalar degrees of
freedom resonate through oscillations at the bottom of their
potential.  During reheating, the universe evolves effectively as a
dust-dominated one, much as it does in the SFDM model, but at
significantly higher energy scales, and thus covering a different sector of the parameter-space. 
Both stages, however, are often modeled with a real or complex scalar field, subject to a quadratic (free-field) potential.  

SFDM matches most of cold dark matter predictions, with the advantage
of solving some of CDM difficulties (for a comprehensive review
  see Ref.~\cite{hui2017ultralight}). Of particular relevance in the SFDM model
is supermassive black hole formation at galactic nuclei, attributed to
the gravitational collapse of the dark matter core at the center of
halos \citep{supermassive,sm7,chavanis2019predictive,mipaper}. In this
letter we propose that the analogous process in the context of
reheating can serve as a mechanism for the production of primordial
black holes (PBHs).

During reheating, primordial {virialized} structures, analogous to
those of the SFDM model, appear if oscillations last long enough and
give way to a ``primordial structure formation" process
\citep{sfdmrh1}, in analogy with the corresponding process in SFDM.
One can thus consider that the reheating process takes place in a
universe filled with inhomogeneities, as a result of the fragmentation
of the inflaton and formation of inflaton halos (inflaton clusters)
and inflaton stars (halo cores) \cite{sfdmrh2,Eggemeier:2020zeg}.

In the present work we explore the possibility of PBH formation via
the collapse of structures formed as a result of the virialization of
scalar field inhomogeneities. We study two formation scenarios for
PBHs: we first determine the conditions for the complete virialized
structure to lie within its Schwarzschild radius, which corresponds to
the gravitational collapse of the inflaton halos. Secondly, we present the requirements for the collapse of the central region of the inflaton halos, the inflaton stars.

We find that for the free-field, the density contrast required for the inflaton star collapse is an order of magnitude smaller than the threshold value required by the collapse of the complete halo. This in part is because,  as we show, the collapsed halos result in black holes about 50 times more massive than the collapsed inflaton stars. We present an extension of these results for the case of a self-interacting scalar field, and finally discuss the implications for the abundance of PBHs formed during reheating, beyond the idealized dust environment. 

\section{The oscillatory field during Reheating} 
Once the accelerated expansion ceases in the early universe, it is
assumed that the inflaton rolls down quickly to the minimum of its
potential and oscillates rapidly about the minimum. In this regime, and for
a large variety of inflationary potentials, the fast oscillations of
the (real or complex) scalar field are captured by the first term of a
Taylor expansion of the potential around its minimum, the harmonic
potential
\begin{equation}
V(\phi) = \frac{\mu^2}{2} |\phi|^2\,,
\label{potential:phi}
\end{equation}
where the (effective) mass of the scalar field $\mu$ dictates the
characteristic oscillation frequency\footnote{We take the potential \eqref{potential:phi} as valid in the post-inflationary universe, with {$\mu$} as a free parameter. The field $\phi$ in question may or may not be the inflaton itself.}.  The fast oscillations regime is
reached when the Hubble scale falls well below the mass scale ($H \ll
\mu$). Averaged over a Hubble time, both real and complex scalar
fields present similar dynamics. We thus focus on the solutions for
the complex field, which require no averaging of the fast oscillation
in the cosmological quantities. In other words, our results are valid
for both complex and real fields after averaging.  For the particular
potential \eqref{potential:phi}, the evolution of the scalar field is
such that the background energy density scales like a dust component, {which for a  flat Friedmann-Robertson-Walker background is given by}
\begin{equation}\label{rho_0_m}
  \rho_0(a) = \frac{3m_{\rm Pl}^2H_{\rm end}^2}{8\pi}
  \left(\frac{a_{\rm end}}{a}\right)^3\,,
\end{equation}
where the subscript ${}_{\rm end}$ refers to quantities evaluated at the end of inflation, $a$ is the scale factor, {and $m_{\rm Pl}$ is the Planck mass}. 

\section{Structure in the early universe}
In the canonical mechanism, the $k$--modes associated with the quantum fluctuations of the inflaton present a fixed amplitude when stretching beyond the Hubble horizon, during the accelerated expansion phase (thereby suppressing the decaying mode).  After inflation, these modes re-enter the cosmological horizon, giving way to a process of {structure formation}, analogous to that of the SFDM model. 
Inside the cosmological horizon and in the fast oscillating phase, two regimes are distinguished, which are separated by the scale $k_Q$ -- usually referred to as the \textit{Quantum Jeans scale} or simply the \textit{Jeans scale} -- given by~\citep{jeans3}
\begin{equation}
    k_Q = \left({16\pi G \rho_0 \mu^2a^4}\right)^{1/4}.
\end{equation}
Density fluctuations with a characteristic scale $k> k_Q$ {are damped
  through oscillations}, whereas those with a size $k< k_Q$ behave as
standard dust inhomogeneities, which in the perturbative regime grow
in amplitude at the rate of the scale factor. It is well known that
the description of a single dust component can be fully captured
through the Newtonian formalism \citep{Ellis:1971pg}. In fact, it has been argued that the
Newtonian treatment of the system is enough to describe the
post-inflationary dynamics \citep{sfdmrh1}, and we shall follow this
formalism for the rest of this work.

Since this process starts from causal contact at the Hubble horizon reentry, one can express the density contrast as
\begin{equation}\label{delta_a}
    \delta(a;k) = \delta_{\rm HC}(k)\frac{a}{a_{\rm HC}}\,,
\end{equation}
where the subscript ${}_{\rm{HC}}$  refers to quantities evaluated at the horizon crossing time. 

Once detached from the expansion, throughout the nonlinear regime,
overdensities of size much larger than the critical value (in this
case, the de Broglie wavelength $\lambda_{\rm dB}$), present the same
evolution as dust configurations. Indeed, an equivalence has been
drawn between the Sch\"odinger-Poisson system (used to describe the
non-linear evolution of the scalar field at scales smaller than the
cosmological horizon and for non-relativistic velocities) and the
Vlasov-Poisson equations (see for example
Refs.~\cite{Widrow:1993qq,mocz2018schrodinger}).

Overdensities of pressureless dust form virialized structures on a
timescale of order the dynamical time, $t_{\rm NL}\simeq (4\pi G
\rho_0(a_{\rm NL}))^{-1/2}$, where $a_{\rm NL}$ refers to the scale
factor measured at this time. In regard of Eq. \eqref{rho_0_m},
this time is given by
%
  \label{t_din}
\begin{equation}
  t_{\rm NL} = 4.4\times 10^{-39}\left(\frac{10^{-5}m_{\rm Pl}}{H_{\rm end}}\right)
  e^{\frac{3}{2}(N_{\rm HC}(k)+N_{\rm NL}(k))}~\rm{s}\,, 
  \label{t_din}
\end{equation}
where we have expressed $a_{\rm NL}(k) = a_{\rm end}e^{N_{\rm HC}(k)+N_{\rm NL}(k)}$, in terms of the number of $e$-folds required for the inhomogeneity to reenter the cosmological horizon $N_{\rm HC}$, and to become non-linear after horizon crossing, $N_{\rm NL}$. Explicitly, for the overdensity of size $ 2\pi/k$,
\begin{equation}\label{Ns}
    N_{\rm HC}(k) = 2\ln{\left(\frac{k_{\rm end}}{k}\right)}\,,   \ \ \
    N_{\rm NL}(k) = \ln\left(1.39\,\delta_{\rm HC}^{-1}(k)\right)\,.
\end{equation}
Note that from the top-hat spherical collapse model, we have taken the value $\delta_{\rm NL} = 1.39$ for the linear overdensity to identify the virialization time\footnote{In the top-hat model an overdensity should virialize at a radius of order $r_{\rm max}/2$, where $r_{\rm max}$ is the turn around radius. At that time the value of the density contrast predicted by the linear theory is $\delta \simeq 1.39$.}. 

The final result of this process are quasi-spherical halos with a
specific density profile, close to an NFW function. Such structures also arise in numerical simulations
of the SFDM scenario {after averaging over the angular coordinates}
\citep{sc4,sc10}. Simulations of the fast oscillating regime of the
reheating field confirm the analogy with the SFDM model and the
formation of similar, primordial structures, referred to as
\textit{inflaton halos} \cite{sfdmrh2,Eggemeier:2020zeg}.

For a particular $k$--mode, the mass of these inflaton halos can be expressed {as}
%
\begin{equation}\label{M_ih2}
    M_{\rm IH} = \frac{m_{\rm Pl}^2}{2H_{\rm end}}\left(\frac{a_{\rm HC}}{a_{\mathrm{end}}}\right)^{3/2}(1+\delta_{\mathrm{end}}(k)).
\end{equation}
As in SFDM structure formation, we take $M_{\rm IH} = 4\pi R_{\rm IH}^3 \rho_{200}(a_{\rm NL})/3$, where $\rho_{200}(a_{\rm NL}) =  200\rho_0(a_{\rm NL})$ and $R_{\rm IH}$ is the virial radius of the inflaton halo, we find that these structures should virialize and halt collapse at radius
\begin{equation}\label{R_vir}
  R_{\rm IH}\simeq
  \left(\frac{3}{4\pi}\frac{M_{\rm IH}}{\rho_{200}(a_{\rm NL})}\right)^{1/3}\,.
\end{equation}
Within these inflaton halos, and on scales close to the de Broglie wavelength, additional phenomena associated to the wave-like dynamics of the scalar field are expected. Particularly, the formation of boson stars (or soliton structures) through Bose-Einstein condensation takes place in the center of inflaton halos. {In the context of reheating, these core condensations are called} \textit{inflaton stars} \cite{sfdmrh2,Eggemeier:2020zeg}.
The condensation time $\tau$ required for these structures to form can be derived from the theory of relaxation in the Schr\"odinger-Poisson equations and the Landau equation (see Ref.~\citep{sc14} for details),
\begin{equation}
  \tau=8.168\times 10^{-18}\left(\mu_5^2M_{\rm IH}R_{\rm IH}\right)^{3/2}t_{\rm NL}\,,
  \label{t_cond}
\end{equation}
{ where $\mu_5\equiv \mu/(10^{-5}m_{\rm Pl})$}.

In the SFDM model, there is {a specific} relation between the total
mass of the soliton structure in the center of halos and the total
mass of the complete halo. We can derive such a relation through
various criteria: either, by demanding that the velocity dispersion
at the radius of the soliton be the same as that at the edge of the
halo; or by demanding that the core and the halo present the same virial
temperature; or by demanding an equivalence between the energy/mass
ratio in the soliton and in the complete halo (see for example
Refs.~\cite{chavanis2019predictive,mipaper}). In all cases the
proportionality $M_{\rm IH}/R_{\rm IH}\simeq M_{\rm IS}/R_{\rm IS}$
between inflaton star (IS) and the complete inflaton halo (IH) is
observed. On the other hand, the soliton profiles fulfill the
mass-radius relation $M_{\rm IS} = (9.9 m_{\rm Pl}^2/\mu)
\cdot(\mu^{-1}/R_{\rm IH})$, where we combine all these expressions,
to arrive at
{\begin{equation}
    \left( \frac{M_{\rm IS}}{10^{20}~\rm{GeV}}\right) = 8.61\frac{\hat\rho^{1/6}_{11}(a_{\rm NL})}{ \mu_5}\left(\frac{M_{\rm IH}}{10^{24}~\rm{GeV}}\right)^{1/3}\,,
    \label{mcmhrel}
\end{equation}}
where 
{$\hat\rho_{11}(a_{\rm NL})\equiv \rho_{200}(a_{\rm NL})/(10^{11}~\rm{GeV})^4$}.

\section{Primordial black hole formation}
\subsection {Case 1: Inflaton halo collapse}
If the structures that form after inflation are very massive, one
might expect that they could be gravitationally unstable and collapse
forming a PBH on a time scale close to the dynamical time \eqref{t_din}. In
particular, from the description above, a good indicator of such
collapse is whether the overdensities virialize at a radius similar or
smaller than the Schwarzschild radius {$R_{\rm Sch}\equiv 2GM_{\rm
    IH}$} associated to the halo, i.e. if $R_{\rm IH}\leq R_{\rm
  Sch}$. This condition implies from Eq. \eqref{R_vir} that
{\begin{equation}
    M_{\rm IH}\geq \frac{3.144\times 10^{34}}{\sqrt{\hat \rho_{11}(a_{\rm NL})}}~\rm{GeV}\,.
    \label{M_ih_col}
\end{equation}}
When combined with Eqs. \eqref{rho_0_m} and \eqref{M_ih2}, this inequality can be expressed in terms of the density contrast at the end of inflation. After some algebra we arrive at
\begin{equation}
  \label{N_nl_col}
    N_{\rm NL}\leq \frac{2}{3}\ln[14.14(1+\delta_{\rm end})]\,.
\end{equation}
Using Eq. \eqref{Ns} we can turn this inequality into a condition for the amplitude of the density contrast at the horizon crossing time to form a primordial black hole:
\begin{equation}
  \label{dnl_tot}
    \delta_{\rm HC} \geq  \frac{1.39}{[14.14(1+\delta_{\rm end})]^{2/3}}\,.
\end{equation}
Note that in the limit $\delta_{\rm end}\ll 1$, which is a typical value for most inflationary models, the collapse occurs when 
\begin{equation}\label{Nnl_tot}
    N_{\rm NL}\leq 1.766, \quad \mathrm{and} \quad \delta_{\rm HC} \geq \delta_{\rm crit}^{(\rm IH)}\equiv 0.238. 
\end{equation}

\subsection{Case 2: Inflaton star collapse}
Inflaton stars are understood as configurations which balance the
attractive force of gravity, and the quantum repulsion associated to
the Heisenberg uncertainty principle (see
e.g.~Ref.~\cite{Schunck:2003kk}). However, if a soliton generated in
this scenario is massive enough, it is expected that the quantum
repulsion cannot balance self-gravity, and the structure may collapse
to form a black hole. This implies the existence of a maximum mass for
which inflaton stars can be stable, a phenomenon equivalent to the
Chandrasekhar mass limit for white dwarf stars but associated to
soliton structures. The numerical value of this critical mass is
$M_{\rm IS}^{\rm crit} = 0.633 m_{\rm Pl}^2/\mu$
\citep{boson_stars1,osc1,gc1}. If we substitute this value into
Eq.~\eqref{mcmhrel}, we find a critical halo mass $M_{\rm IH}^{\rm
  crit}$ beyond which its central soliton should be unstable and to
form a PBH in a time close to the condensation time
\eqref{t_cond}. Here the condition for PBH formation is given by
\begin{equation}
    M_{\rm IH}\geq
    \frac{7.237\times 10^{32}}{\sqrt{\hat \rho_{11}(a_{\rm NL})}}~\rm{GeV}\,.
    \label{M_crit_free}
\end{equation}
Comparing relation (\ref{M_crit_free}) with Eq. \eqref{M_ih_col}, we infer that
the soliton collapse may be reached for smaller halo masses than for
the case of complete halo collapse, and therefore may be achieved from
a smaller density contrast. Repeating the above algebra we arrive at
the corresponding threshold values for PBH formation
\begin{equation}
  N_{\rm NL}^{\rm (free)}\leq 4.28,  \quad \mathrm{and} \quad \delta_{\rm HC}\geq \delta_{\rm crit}^{\rm (free)}\equiv 0.019\,.
\label{Nnl_tot2}
\end{equation}

\section{A simple extension: The self-interacting scenario} 
Alternatives to the free-field potential \eqref{potential:phi} are also considered in the reheating scenario. {The stable structures formed in each model show a singular mass-to-radius relation   \citep{chavanis2019predictive,mipaper}, thereby modifying the collapse threshold values of PBH formation, as we are about to show}. We exemplify this by looking at the self-interaction model, which is of wide interest in the SFDM scenario. In this case, the potential can be expressed as
\begin{equation}
    V(\phi) = \frac{1}{2}{\mu^2 }|\phi|^2+\frac{\lambda}{4 }|\phi|^4\,,
\end{equation}
where $\lambda$ parametrizes the self-interaction of the boson particles that constitute the reheating field.
This self-interaction is often regarded as a pressure force that can be repulsive or attractive, if {$\lambda$} is positive or negative, respectively. {Moreover, this potential can be interpreted as the leading order terms in the Taylor expansion of a more complicated inflationary, symmetric potential (ignoring the constant term, as required in reheating).} Our focus here is the attractive scenario, since it would help the formation of PBHs.   

At the background level, a cosmological solution for the scalar field with an attractive self-interaction eventually matches the pressureless dust evolution of Eq.~\eqref{rho_0_m} \citep{charge4}. Moreover, at the perturbative order, the incorporation of the self-interacting term could cause an exponential growth of the density contrast at small scales, which could enhance the formation of primordial structures \citep{jeans3}. This is because the self-interaction of the particles defines a new characteristic scale in the system, given by: 
%
\begin{equation}
\label{self-int:k}
    k_{\rm SI} = \sqrt{8 \pi \left(\frac{\mu^2}{|\lambda|}\right)}\left(\frac{\mu}{m_{\rm Pl}}\right)a\,.
\end{equation}
On the other hand, for large enough scales, which are of primary interest here, the evolution mimics that of a dust-like fluid. This implies that, after horizon crossing, inhomogeneities are described by Eq.~\eqref{delta_a} and, in consequence, the threshold for the inflaton halo collapse remains invariant. 

In view of the above, the evolution of a self-interacting inflaton halo is equivalent to that in the free field scenario. This is not the case for inflaton stars, which are expected to show a modified star-to-halo mass relation. 
Such modification was explored in great detail in \citep{chavanis2019predictive,mipaper} for SFDM models, so here we briefly show how to adapt it to the reheating scenario. 

In extension to the free case, the mass-radius relationship of the inflaton star is now given by $M_{\rm IS} = 9.9(m_{\rm Pl}^2/\mu)(\mu^{-1}/R_{\rm IS})/[1-3\pi^2\Lambda(\mu^{-1}/R_{\rm IS})^2]$, with $\Lambda\equiv  (\lambda/8\pi) (m_{\rm Pl}/\mu)^2$. This implies that the relation in Eq.~\eqref{mcmhrel} is now modified to
{\begin{eqnarray}
    &&\left( \frac{M_{\rm IS}}{10^{20}~\rm{GeV}}\right) = 3.43\frac{\hat\rho^{1/6}_{11}(a_{\rm NL})}{ \mu_5}\left(\frac{M_{\rm IH}}{10^{24}~\rm{GeV}}\right)^{1/3}\times \nonumber\\
    &&\times \sqrt{1-59.73|\lambda|\frac{\hat\rho_{11}^{1/3}}{\mu_5^2}\left(\frac{M_{\rm IH}}{10^{24}~\rm{GeV}}\right)^{2/3}}\,.
\end{eqnarray}}
From this expression, 
{$M_{\rm IS}^{\rm (crit)} = (5.57\times 10^{19}/\sqrt{|\lambda|}) ~\rm{GeV}$} is a maximum of the inflaton star mass.
For larger masses a PBH should form at the core. In turn, this will occur when halo masses fulfill the condition 
\begin{equation}\label{M_crit_self}
    M_{\rm IH}\geq 
    \frac{2.166\times 10^{21}}{\sqrt{\hat\rho_{11}(a_{\rm NL})}}\left(\frac{|\lambda|}{\mu_5^2}\right)^{-3/2}~\rm{GeV}\,.
\end{equation}
If we follow the same procedure as above, which led to Eqs.~\eqref{Nnl_tot} and \eqref{Nnl_tot2}, {we} arrive at the conditions
\begin{subequations}
\begin{equation}
    N_{\rm NL} \leq \frac{2}{3}\ln\left[5.804\times 10^{13}\left(\frac{|\lambda|}{\mu_5^2}\right)^{3/2}\right],
\end{equation}
\begin{equation} \label{Nnl_tot3}
  \mathrm{and}\quad  \delta_{\rm HC}\geq \delta_{\rm crit}^{(\rm SI)}\equiv {9.27\times 10^{-10}}{\left(\frac{|\lambda|}{\mu_5^2}\right)^{-1}}\,.
\end{equation}
\end{subequations}

We emphasize that the collapse in this model will not occur for all values of $\lambda$. The restriction for this parameter comes from demanding that the critical mass of collapse \eqref{M_crit_self} be smaller than that of the free case \eqref{M_crit_free}. This is guaranteed for values 
\begin{equation}\label{si_col_cond}
    \frac{|\lambda|}{\mu_5^2}\geq 4.88\times 10^{-8}\,.
\end{equation}

\section{Probability of primordial black hole formation} 
%
To illustrate the relevance of our results, we compute the probability of primordial black hole formation in light of the featured mechanism, together with previous estimates relevant to the reheating phase. Considering Gaussian density perturbations with variance $\sigma^2$, the fraction of horizon-size patches collapsing onto PBHs of a single mass (monochromatic) is given by \citep{carr1975primordial}
\begin{equation}
\label{beta:delta_c}
    \beta\simeq \mathrm{Erfc}\left[\frac{\delta_\mathrm{\rm crit}}{\sqrt{2}\sigma}\right]\,, 
\end{equation}
where $\mathrm{Erfc}(x)$ is the complementary error function and $\delta_{\rm crit}$ is the threshold value of the density contrast for PBH formation, evaluated at horizon crossing. In a universe dominated by radiation, this threshold value is $\delta_{\rm crit}\simeq~0.41$ \citep{Musco:2012au}, whereas for reheating we find that such values are given by Eq.~\eqref{Nnl_tot} in the case of the total inflaton halo collapse, Eq.~\eqref{Nnl_tot2} in the case of the inflaton star collapse in the free-field case, and Eq.~\eqref{Nnl_tot3} in the case of the inflaton star collapse with an attractive self-interaction. 

\begin{figure}
\begin{center}
        \includegraphics[width=0.95\linewidth]{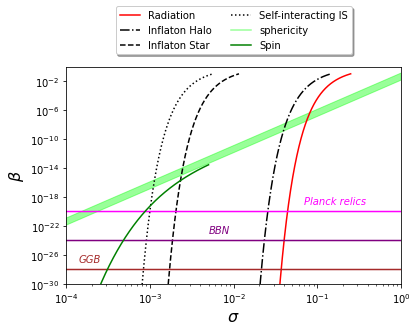}
    \captionof{figure}[The PBH density fraction $\beta$]{\small{The PBH density fraction $\beta$ plotted as a
        function of the mean density contrast $\sigma$, as given by
        Eq.~\eqref{beta:delta_c} and evaluated with the threshold
        amplitude derived for each case. We     used $\lambda/\mu_5^2= 10^{-7}$. The
        maximum values of $\beta$ allowed in a dust-dominated universe
        are plotted as two green lines. The thick green line accounts for
        the sphericity condition for PBH formation
        \citep{harada2016primordial}, whereas the thin green curve
        shows the condition for the maximum spin allowed for collapse
        \citep{Harada:2017fjm}. For reference, we include
        observational constraints on $\beta$ \cite{carr2020constraints}, provided by the galactic
        gamma-ray background (brown), big bang nucleosynthesis
        (purple), and Planck mass relics (pink), each of these constraining monochromatic mass functions at the relevant masses.}}
    \label{fig:beta_vs_sigma}
\end{center}
\end{figure}

We show in Fig.~\ref{fig:beta_vs_sigma} the mass fraction $\beta$ as a function of the mean amplitude $\sigma$ for radiation and the three scenarios featured here. The probability of PBH formation is smallest in the radiation era. Moreover, the collapse of the inflaton halo is less likely than the collapse of the inflaton star, mostly due to the mass difference between configurations.
An attractive self-interaction between the particles requires a much smaller variance, as expected, to produce PBHs significantly.

{If primordial perturbations follow a Gaussian distribution, PBHs are expected to form most abundantly with masses close to the critical mass in each case, since the probability of perturbations with higher masses is exponentially suppressed. As an example, taking $\mu_5 = 1$, the PBHs that formed due to the collapse of inflaton stars in the free-field potential present masses $M_{\rm PBH} \approx 10^{-3}~\rm{kg}$, whereas in the self-interacting case with $\lambda/\mu_5^2 =10^{-7}$ such objects present $M_{\rm PBH}\approx 3 \times 10^{-4}~\rm{kg}$. As shown in Fig.~\ref{fig:beta_vs_sigma}, these black holes are subject to the Planck mass relics constraint, which is saturated for primordial fluctuations as small as $\sigma = 0.002$ and $\sigma = 0.001$, respectively, if thermalization follows immediatly after black hole formation.}

For comparison we include in Fig.~\ref{fig:beta_vs_sigma} the constraints on the abundance of PBHs formed in a dust-like background (see, for details Refs.~\cite{harada2016primordial,Harada:2017fjm}).
%
%
%
%
Relevant constraints due to the bounds on PBH formation are represented by horizontal lines \cite{carr2020constraints}. Our figure shows how likely it is to form PBHs in a scalar field-dominated phase in comparison with the
radiation-dominated or dust-dominated universe. The low values of
$\sigma$ required for a sizable $\beta$ in the self-interacting model,
indicate that this model is subject to the same conditions as the
dust-dominated scenario for large enough $\sigma$. Otherwise, {the inflaton star collapse can evade the constraints of a dust scenario and require a smaller $\sigma$ to overproduce PBHs.}

\section{Conclusions and discussion} 
%
We have put forward a new mechanism of PBH formation during single
scalar-field domination, realized in the reheating period. This mechanism
is based on the observation that sufficiently massive nonlinear structures (inflaton halos) could have formed within their Schwarzschild radius, or
could be massive enough to overcome the quantum pressure of the central core (inflaton
stars) and thus form PBHs. These objects are produced more effectively
in a free-field dominated universe than in the radiation era, but require larger overdensities than those allowed to collapse in a matter-dominated
universe (as assumed previously during reheating, e.g. Refs.~\cite{Green:2000he,Hidalgo:2017dfp,Martin:2019nuw,reh5}). 
Considering an attractive self-interaction, 
the probability of PBH formation can saturate the sphericity and rotation conditions that constrain the PBH production in the matter-dominated scenario. These results represent a solid tool to constrain reheating models through the bounds to PBH abundance.

Simulations of scalar field collapse are still to provide accurate conditions for PBH formation during the reheating stage (see however Ref.~\cite{2021arXiv210904896D}). Our novel mechanism provides explicit conditions for PBH formation beyond the simplistic dust-like scenario. {The precise description of PBH formation is crucial since these objects constitute an observable \cite{carr2020constraints}. {Specifically, in the lower mass end of the spectrum, the possible Planck mass relics left behind after the evaporation of small black holes may constitute part or all of dark matter \citep{Martin:2019nuw,Carrion:2021yeh}. Additionally, a stochastic gravitational wave background may be generated due to the collapse of virialized structures \cite{Jedamzik:2010hq,Dalianis:2020gup}, and from the collision of primordial black holes \cite{Sasaki:2018dmp}. In particular, for the mechanism put forward in this paper, the delay in the process of formation with respect to the standard case results in a delayed evaporation and the associated gravitational waves produced at a later times. The impact of these effects in observables and constraints will be reported elsewhere. This information,} in tandem with information of the primordial spectrum, can provide clues on the physics of the reheating phase.} \\

\begin{acknowledgments}
{This work is sponsored by CONACyT Network Project No.~304001 ``Estudio
de campos escalares con aplicaciones en cosmolog\'ia y
astrof\'isica''. LEP and JCH acknowledge sponsorship from CONACyT
through grant CB-2016-282569 and from program UNAM-PAPIIT, grant
IN107521 ``Sector Oscuro y Agujeros Negros Primordiales''. The work of
LEP is also supported by the DGAPA-UNAM postdoctoral grants
program. KAM is supported in part by STFC grant ST/T000341/1.}
\end{acknowledgments}
\bibliographystyle{apsrev4-1}
\bibliography{references}

\begin{thebibliography}{40}%
\makeatletter
\providecommand \@ifxundefined [1]{%
 \@ifx{#1\undefined}
}%
\providecommand \@ifnum [1]{%
 \ifnum #1\expandafter \@firstoftwo
 \else \expandafter \@secondoftwo
 \fi
}%
\providecommand \@ifx [1]{%
 \ifx #1\expandafter \@firstoftwo
 \else \expandafter \@secondoftwo
 \fi
}%
\providecommand \natexlab [1]{#1}%
\providecommand \enquote  [1]{``#1''}%
\providecommand \bibnamefont  [1]{#1}%
\providecommand \bibfnamefont [1]{#1}%
\providecommand \citenamefont [1]{#1}%
\providecommand \href@noop [0]{\@secondoftwo}%
\providecommand \href [0]{\begingroup \@sanitize@url \@href}%
\providecommand \@href[1]{\@@startlink{#1}\@@href}%
\providecommand \@@href[1]{\endgroup#1\@@endlink}%
\providecommand \@sanitize@url [0]{\catcode `\\12\catcode `\$12\catcode
  `\&12\catcode `\#12\catcode `\^12\catcode `\_12\catcode `\%12\relax}%
\providecommand \@@startlink[1]{}%
\providecommand \@@endlink[0]{}%
\providecommand \url  [0]{\begingroup\@sanitize@url \@url }%
\providecommand \@url [1]{\endgroup\@href {#1}{\urlprefix }}%
\providecommand \urlprefix  [0]{URL }%
\providecommand \Eprint [0]{\href }%
\providecommand \doibase [0]{http://dx.doi.org/}%
\providecommand \selectlanguage [0]{\@gobble}%
\providecommand \bibinfo  [0]{\@secondoftwo}%
\providecommand \bibfield  [0]{\@secondoftwo}%
\providecommand \translation [1]{[#1]}%
\providecommand \BibitemOpen [0]{}%
\providecommand \bibitemStop [0]{}%
\providecommand \bibitemNoStop [0]{.\EOS\space}%
\providecommand \EOS [0]{\spacefactor3000\relax}%
\providecommand \BibitemShut  [1]{\csname bibitem#1\endcsname}%
\let\auto@bib@innerbib\@empty
\bibitem [{\citenamefont {Matos}\ and\ \citenamefont
  {Urena-Lopez}(2000)}]{Matos:2000ng}%
  \BibitemOpen
  \bibfield  {author} {\bibinfo {author} {\bibfnamefont {T.}~\bibnamefont
  {Matos}}\ and\ \bibinfo {author} {\bibfnamefont {L.}~\bibnamefont
  {Urena-Lopez}},\ }\href {\doibase 10.1088/0264-9381/17/13/101} {\bibfield
  {journal} {\bibinfo  {journal} {Class. Quant. Grav.}\ }\textbf {\bibinfo
  {volume} {17}},\ \bibinfo {pages} {L75} (\bibinfo {year} {2000})},\ \Eprint
  {http://arxiv.org/abs/astro-ph/0004332} {arXiv:astro-ph/0004332} \BibitemShut
  {NoStop}%
\bibitem [{\citenamefont {Hui}\ \emph {et~al.}(2017)\citenamefont {Hui},
  \citenamefont {Ostriker}, \citenamefont {Tremaine},\ and\ \citenamefont
  {Witten}}]{hui2017ultralight}%
  \BibitemOpen
  \bibfield  {author} {\bibinfo {author} {\bibfnamefont {L.}~\bibnamefont
  {Hui}}, \bibinfo {author} {\bibfnamefont {J.~P.}\ \bibnamefont {Ostriker}},
  \bibinfo {author} {\bibfnamefont {S.}~\bibnamefont {Tremaine}}, \ and\
  \bibinfo {author} {\bibfnamefont {E.}~\bibnamefont {Witten}},\ }\href@noop {}
  {\bibfield  {journal} {\bibinfo  {journal} {Physical Review D}\ }\textbf
  {\bibinfo {volume} {95}},\ \bibinfo {pages} {043541} (\bibinfo {year}
  {2017})}\BibitemShut {NoStop}%
\bibitem [{\citenamefont {Tsujikawa}(2013)}]{quintessence}%
  \BibitemOpen
  \bibfield  {author} {\bibinfo {author} {\bibfnamefont {S.}~\bibnamefont
  {Tsujikawa}},\ }\href {\doibase 10.1088/0264-9381/30/21/214003} {\bibfield
  {journal} {\bibinfo  {journal} {Class. Quant. Grav.}\ }\textbf {\bibinfo
  {volume} {30}},\ \bibinfo {pages} {214003} (\bibinfo {year} {2013})},\
  \Eprint {http://arxiv.org/abs/1304.1961} {arXiv:1304.1961 [gr-qc]}
  \BibitemShut {NoStop}%
\bibitem [{\citenamefont {Guth}(1981)}]{Guth}%
  \BibitemOpen
  \bibfield  {author} {\bibinfo {author} {\bibfnamefont {A.~H.}\ \bibnamefont
  {Guth}},\ }\href@noop {} {\bibfield  {journal} {\bibinfo  {journal} {Physical
  Review D}\ }\textbf {\bibinfo {volume} {23}},\ \bibinfo {pages} {347}
  (\bibinfo {year} {1981})}\BibitemShut {NoStop}%
\bibitem [{\citenamefont {{Linde}}(1982)}]{Linde1982}%
  \BibitemOpen
  \bibfield  {author} {\bibinfo {author} {\bibfnamefont {A.~D.}\ \bibnamefont
  {{Linde}}},\ }\href {\doibase 10.1016/0370-2693(82)91219-9} {\bibfield
  {journal} {\bibinfo  {journal} {Physics Letters B}\ }\textbf {\bibinfo
  {volume} {108}},\ \bibinfo {pages} {389} (\bibinfo {year}
  {1982})}\BibitemShut {NoStop}%
\bibitem [{\citenamefont {Allahverdi}\ \emph {et~al.}(2010)\citenamefont
  {Allahverdi}, \citenamefont {Brandenberger}, \citenamefont {Cyr-Racine},\
  and\ \citenamefont {Mazumdar}}]{reh2}%
  \BibitemOpen
  \bibfield  {author} {\bibinfo {author} {\bibfnamefont {R.}~\bibnamefont
  {Allahverdi}}, \bibinfo {author} {\bibfnamefont {R.}~\bibnamefont
  {Brandenberger}}, \bibinfo {author} {\bibfnamefont {F.-Y.}\ \bibnamefont
  {Cyr-Racine}}, \ and\ \bibinfo {author} {\bibfnamefont {A.}~\bibnamefont
  {Mazumdar}},\ }\href {\doibase 10.1146/annurev.nucl.012809.104511} {\bibfield
   {journal} {\bibinfo  {journal} {Ann. Rev. Nucl. Part. Sci.}\ }\textbf
  {\bibinfo {volume} {60}},\ \bibinfo {pages} {27} (\bibinfo {year} {2010})},\
  \Eprint {http://arxiv.org/abs/1001.2600} {arXiv:1001.2600 [hep-th]}
  \BibitemShut {NoStop}%
\bibitem [{\citenamefont {Lozanov}(2019)}]{Lozanov:2019jxc}%
  \BibitemOpen
  \bibfield  {author} {\bibinfo {author} {\bibfnamefont {K.~D.}\ \bibnamefont
  {Lozanov}},\ }\href@noop {} {\  (\bibinfo {year} {2019})},\ \Eprint
  {http://arxiv.org/abs/1907.04402} {arXiv:1907.04402 [astro-ph.CO]}
  \BibitemShut {NoStop}%
\bibitem [{\citenamefont {Chavanis}(2016)}]{supermassive}%
  \BibitemOpen
  \bibfield  {author} {\bibinfo {author} {\bibfnamefont {P.-H.}\ \bibnamefont
  {Chavanis}},\ }\href@noop {} {\bibfield  {journal} {\bibinfo  {journal}
  {Physical Review D}\ }\textbf {\bibinfo {volume} {94}},\ \bibinfo {pages}
  {083007} (\bibinfo {year} {2016})}\BibitemShut {NoStop}%
\bibitem [{\citenamefont {\'Avilez}\ \emph {et~al.}(2018)\citenamefont
  {\'Avilez}, \citenamefont {Bernal}, \citenamefont {Padilla},\ and\
  \citenamefont {Matos}}]{sm7}%
  \BibitemOpen
  \bibfield  {author} {\bibinfo {author} {\bibfnamefont {A.}~\bibnamefont
  {\'Avilez}}, \bibinfo {author} {\bibfnamefont {T.}~\bibnamefont {Bernal}},
  \bibinfo {author} {\bibfnamefont {L.~E.}\ \bibnamefont {Padilla}}, \ and\
  \bibinfo {author} {\bibfnamefont {T.}~\bibnamefont {Matos}},\ }\href@noop {}
  {\bibfield  {journal} {\bibinfo  {journal} {Monthly Notices of the Royal
  Astronomical Society}\ }\textbf {\bibinfo {volume} {477}},\ \bibinfo {pages}
  {3257} (\bibinfo {year} {2018})}\BibitemShut {NoStop}%
\bibitem [{\citenamefont {Chavanis}(2019)}]{chavanis2019predictive}%
  \BibitemOpen
  \bibfield  {author} {\bibinfo {author} {\bibfnamefont {P.-H.}\ \bibnamefont
  {Chavanis}},\ }\href {\doibase 10.1103/PhysRevD.100.123506} {\bibfield
  {journal} {\bibinfo  {journal} {Phys. Rev. D}\ }\textbf {\bibinfo {volume}
  {100}},\ \bibinfo {pages} {123506} (\bibinfo {year} {2019})},\ \Eprint
  {http://arxiv.org/abs/1905.08137} {arXiv:1905.08137 [astro-ph.CO]}
  \BibitemShut {NoStop}%
\bibitem [{\citenamefont {Padilla}\ \emph {et~al.}(2021)\citenamefont
  {Padilla}, \citenamefont {Rindler-Daller}, \citenamefont {Shapiro},
  \citenamefont {Matos},\ and\ \citenamefont {Alberto~V\'azquez}}]{mipaper}%
  \BibitemOpen
  \bibfield  {author} {\bibinfo {author} {\bibfnamefont {L.~E.}\ \bibnamefont
  {Padilla}}, \bibinfo {author} {\bibfnamefont {T.}~\bibnamefont
  {Rindler-Daller}}, \bibinfo {author} {\bibfnamefont {P.~R.}\ \bibnamefont
  {Shapiro}}, \bibinfo {author} {\bibfnamefont {T.}~\bibnamefont {Matos}}, \
  and\ \bibinfo {author} {\bibfnamefont {J.}~\bibnamefont
  {Alberto~V\'azquez}},\ }\href {\doibase 10.1103/PhysRevD.103.063012}
  {\bibfield  {journal} {\bibinfo  {journal} {Phys. Rev. D}\ }\textbf {\bibinfo
  {volume} {103}},\ \bibinfo {pages} {063012} (\bibinfo {year} {2021})},\
  \Eprint {http://arxiv.org/abs/2010.12716} {arXiv:2010.12716 [astro-ph.GA]}
  \BibitemShut {NoStop}%
\bibitem [{\citenamefont {Musoke}\ \emph {et~al.}(2020)\citenamefont {Musoke},
  \citenamefont {Hotchkiss},\ and\ \citenamefont {Easther}}]{sfdmrh1}%
  \BibitemOpen
  \bibfield  {author} {\bibinfo {author} {\bibfnamefont {N.}~\bibnamefont
  {Musoke}}, \bibinfo {author} {\bibfnamefont {S.}~\bibnamefont {Hotchkiss}}, \
  and\ \bibinfo {author} {\bibfnamefont {R.}~\bibnamefont {Easther}},\ }\href
  {\doibase 10.1103/PhysRevLett.124.061301} {\bibfield  {journal} {\bibinfo
  {journal} {Phys. Rev. Lett.}\ }\textbf {\bibinfo {volume} {124}},\ \bibinfo
  {pages} {061301} (\bibinfo {year} {2020})},\ \Eprint
  {http://arxiv.org/abs/1909.11678} {arXiv:1909.11678 [astro-ph.CO]}
  \BibitemShut {NoStop}%
\bibitem [{\citenamefont {Niemeyer}\ and\ \citenamefont
  {Easther}(2020)}]{sfdmrh2}%
  \BibitemOpen
  \bibfield  {author} {\bibinfo {author} {\bibfnamefont {J.~C.}\ \bibnamefont
  {Niemeyer}}\ and\ \bibinfo {author} {\bibfnamefont {R.}~\bibnamefont
  {Easther}},\ }\href {\doibase 10.1088/1475-7516/2020/07/030} {\bibfield
  {journal} {\bibinfo  {journal} {JCAP}\ }\textbf {\bibinfo {volume} {07}},\
  \bibinfo {pages} {030} (\bibinfo {year} {2020})},\ \Eprint
  {http://arxiv.org/abs/1911.01661} {arXiv:1911.01661 [astro-ph.CO]}
  \BibitemShut {NoStop}%
\bibitem [{\citenamefont {Eggemeier}\ \emph {et~al.}(2021)\citenamefont
  {Eggemeier}, \citenamefont {Niemeyer},\ and\ \citenamefont
  {Easther}}]{Eggemeier:2020zeg}%
  \BibitemOpen
  \bibfield  {author} {\bibinfo {author} {\bibfnamefont {B.}~\bibnamefont
  {Eggemeier}}, \bibinfo {author} {\bibfnamefont {J.~C.}\ \bibnamefont
  {Niemeyer}}, \ and\ \bibinfo {author} {\bibfnamefont {R.}~\bibnamefont
  {Easther}},\ }\href {\doibase 10.1103/PhysRevD.103.063525} {\bibfield
  {journal} {\bibinfo  {journal} {Phys. Rev. D}\ }\textbf {\bibinfo {volume}
  {103}},\ \bibinfo {pages} {063525} (\bibinfo {year} {2021})}\BibitemShut
  {NoStop}%
\bibitem [{\citenamefont {Su{\'a}rez}\ and\ \citenamefont
  {Chavanis}(2015)}]{jeans3}%
  \BibitemOpen
  \bibfield  {author} {\bibinfo {author} {\bibfnamefont {A.}~\bibnamefont
  {Su{\'a}rez}}\ and\ \bibinfo {author} {\bibfnamefont {P.-H.}\ \bibnamefont
  {Chavanis}},\ }\href@noop {} {\bibfield  {journal} {\bibinfo  {journal}
  {Physical Review D}\ }\textbf {\bibinfo {volume} {92}},\ \bibinfo {pages}
  {023510} (\bibinfo {year} {2015})}\BibitemShut {NoStop}%
\bibitem [{\citenamefont {Ellis}(1971)}]{Ellis:1971pg}%
  \BibitemOpen
  \bibfield  {author} {\bibinfo {author} {\bibfnamefont {G.~F.~R.}\
  \bibnamefont {Ellis}},\ }\href {\doibase 10.1007/s10714-009-0760-7}
  {\bibfield  {journal} {\bibinfo  {journal} {Proc. Int. Sch. Phys. Fermi}\
  }\textbf {\bibinfo {volume} {47}},\ \bibinfo {pages} {104} (\bibinfo {year}
  {1971})}\BibitemShut {NoStop}%
\bibitem [{\citenamefont {Widrow}\ and\ \citenamefont
  {Kaiser}(1993)}]{Widrow:1993qq}%
  \BibitemOpen
  \bibfield  {author} {\bibinfo {author} {\bibfnamefont {L.~M.}\ \bibnamefont
  {Widrow}}\ and\ \bibinfo {author} {\bibfnamefont {N.}~\bibnamefont
  {Kaiser}},\ }\href@noop {} {\bibfield  {journal} {\bibinfo  {journal}
  {Astrophys. J. Lett.}\ }\textbf {\bibinfo {volume} {416}},\ \bibinfo {pages}
  {L71} (\bibinfo {year} {1993})}\BibitemShut {NoStop}%
\bibitem [{\citenamefont {Mocz}\ \emph {et~al.}(2018)\citenamefont {Mocz},
  \citenamefont {Lancaster}, \citenamefont {Fialkov}, \citenamefont {Becerra},\
  and\ \citenamefont {Chavanis}}]{mocz2018schrodinger}%
  \BibitemOpen
  \bibfield  {author} {\bibinfo {author} {\bibfnamefont {P.}~\bibnamefont
  {Mocz}}, \bibinfo {author} {\bibfnamefont {L.}~\bibnamefont {Lancaster}},
  \bibinfo {author} {\bibfnamefont {A.}~\bibnamefont {Fialkov}}, \bibinfo
  {author} {\bibfnamefont {F.}~\bibnamefont {Becerra}}, \ and\ \bibinfo
  {author} {\bibfnamefont {P.-H.}\ \bibnamefont {Chavanis}},\ }\href@noop {}
  {\bibfield  {journal} {\bibinfo  {journal} {Physical Review D}\ }\textbf
  {\bibinfo {volume} {97}},\ \bibinfo {pages} {083519} (\bibinfo {year}
  {2018})}\BibitemShut {NoStop}%
\bibitem [{\citenamefont {Schive}\ \emph
  {et~al.}(2014{\natexlab{a}})\citenamefont {Schive}, \citenamefont {Chiueh},\
  and\ \citenamefont {Broadhurst}}]{sc4}%
  \BibitemOpen
  \bibfield  {author} {\bibinfo {author} {\bibfnamefont {H.-Y.}\ \bibnamefont
  {Schive}}, \bibinfo {author} {\bibfnamefont {T.}~\bibnamefont {Chiueh}}, \
  and\ \bibinfo {author} {\bibfnamefont {T.}~\bibnamefont {Broadhurst}},\
  }\href@noop {} {\bibfield  {journal} {\bibinfo  {journal} {Nature Physics}\
  }\textbf {\bibinfo {volume} {10}},\ \bibinfo {pages} {496} (\bibinfo {year}
  {2014}{\natexlab{a}})}\BibitemShut {NoStop}%
\bibitem [{\citenamefont {Schive}\ \emph
  {et~al.}(2014{\natexlab{b}})\citenamefont {Schive}, \citenamefont {Liao},
  \citenamefont {Woo}, \citenamefont {Wong}, \citenamefont {Chiueh},
  \citenamefont {Broadhurst},\ and\ \citenamefont {Hwang}}]{sc10}%
  \BibitemOpen
  \bibfield  {author} {\bibinfo {author} {\bibfnamefont {H.-Y.}\ \bibnamefont
  {Schive}}, \bibinfo {author} {\bibfnamefont {M.-H.}\ \bibnamefont {Liao}},
  \bibinfo {author} {\bibfnamefont {T.-P.}\ \bibnamefont {Woo}}, \bibinfo
  {author} {\bibfnamefont {S.-K.}\ \bibnamefont {Wong}}, \bibinfo {author}
  {\bibfnamefont {T.}~\bibnamefont {Chiueh}}, \bibinfo {author} {\bibfnamefont
  {T.}~\bibnamefont {Broadhurst}}, \ and\ \bibinfo {author} {\bibfnamefont
  {W.~P.}\ \bibnamefont {Hwang}},\ }\href@noop {} {\bibfield  {journal}
  {\bibinfo  {journal} {Physical review letters}\ }\textbf {\bibinfo {volume}
  {113}},\ \bibinfo {pages} {261302} (\bibinfo {year}
  {2014}{\natexlab{b}})}\BibitemShut {NoStop}%
\bibitem [{\citenamefont {Levkov}\ \emph {et~al.}(2018)\citenamefont {Levkov},
  \citenamefont {Panin},\ and\ \citenamefont {Tkachev}}]{sc14}%
  \BibitemOpen
  \bibfield  {author} {\bibinfo {author} {\bibfnamefont {D.}~\bibnamefont
  {Levkov}}, \bibinfo {author} {\bibfnamefont {A.}~\bibnamefont {Panin}}, \
  and\ \bibinfo {author} {\bibfnamefont {I.}~\bibnamefont {Tkachev}},\
  }\href@noop {} {\bibfield  {journal} {\bibinfo  {journal} {Physical review
  letters}\ }\textbf {\bibinfo {volume} {121}},\ \bibinfo {pages} {151301}
  (\bibinfo {year} {2018})}\BibitemShut {NoStop}%
\bibitem [{\citenamefont {Schunck}\ and\ \citenamefont
  {Mielke}(2003)}]{Schunck:2003kk}%
  \BibitemOpen
  \bibfield  {author} {\bibinfo {author} {\bibfnamefont {F.~E.}\ \bibnamefont
  {Schunck}}\ and\ \bibinfo {author} {\bibfnamefont {E.~W.}\ \bibnamefont
  {Mielke}},\ }\href {\doibase 10.1088/0264-9381/20/20/201} {\bibfield
  {journal} {\bibinfo  {journal} {Class. Quant. Grav.}\ }\textbf {\bibinfo
  {volume} {20}},\ \bibinfo {pages} {R301} (\bibinfo {year} {2003})},\ \Eprint
  {http://arxiv.org/abs/0801.0307} {arXiv:0801.0307 [astro-ph]} \BibitemShut
  {NoStop}%
\bibitem [{\citenamefont {Seidel}\ and\ \citenamefont
  {Suen}(1990)}]{boson_stars1}%
  \BibitemOpen
  \bibfield  {author} {\bibinfo {author} {\bibfnamefont {E.}~\bibnamefont
  {Seidel}}\ and\ \bibinfo {author} {\bibfnamefont {W.-M.}\ \bibnamefont
  {Suen}},\ }\href@noop {} {\bibfield  {journal} {\bibinfo  {journal} {Physical
  Review D}\ }\textbf {\bibinfo {volume} {42}},\ \bibinfo {pages} {384}
  (\bibinfo {year} {1990})}\BibitemShut {NoStop}%
\bibitem [{\citenamefont {Seidel}\ and\ \citenamefont {Suen}(1991)}]{osc1}%
  \BibitemOpen
  \bibfield  {author} {\bibinfo {author} {\bibfnamefont {E.}~\bibnamefont
  {Seidel}}\ and\ \bibinfo {author} {\bibfnamefont {W.}~\bibnamefont {Suen}},\
  }\href {\doibase 10.1103/PhysRevLett.66.1659} {\bibfield  {journal} {\bibinfo
   {journal} {Phys. Rev. Lett.}\ }\textbf {\bibinfo {volume} {66}},\ \bibinfo
  {pages} {1659} (\bibinfo {year} {1991})}\BibitemShut {NoStop}%
\bibitem [{\citenamefont {Seidel}\ and\ \citenamefont {Suen}(1994)}]{gc1}%
  \BibitemOpen
  \bibfield  {author} {\bibinfo {author} {\bibfnamefont {E.}~\bibnamefont
  {Seidel}}\ and\ \bibinfo {author} {\bibfnamefont {W.-M.}\ \bibnamefont
  {Suen}},\ }\href@noop {} {\bibfield  {journal} {\bibinfo  {journal} {Physical
  review letters}\ }\textbf {\bibinfo {volume} {72}},\ \bibinfo {pages} {2516}
  (\bibinfo {year} {1994})}\BibitemShut {NoStop}%
\bibitem [{\citenamefont {Su{\'a}rez}\ and\ \citenamefont
  {Chavanis}(2017)}]{charge4}%
  \BibitemOpen
  \bibfield  {author} {\bibinfo {author} {\bibfnamefont {A.}~\bibnamefont
  {Su{\'a}rez}}\ and\ \bibinfo {author} {\bibfnamefont {P.-H.}\ \bibnamefont
  {Chavanis}},\ }\href@noop {} {\bibfield  {journal} {\bibinfo  {journal}
  {Physical Review D}\ }\textbf {\bibinfo {volume} {95}},\ \bibinfo {pages}
  {063515} (\bibinfo {year} {2017})}\BibitemShut {NoStop}%
\bibitem [{\citenamefont {Carr}(1975)}]{carr1975primordial}%
  \BibitemOpen
  \bibfield  {author} {\bibinfo {author} {\bibfnamefont {B.~J.}\ \bibnamefont
  {Carr}},\ }\href {\doibase 10.1086/153853} {\bibfield  {journal} {\bibinfo
  {journal} {Astrophys. J.}\ }\textbf {\bibinfo {volume} {201}},\ \bibinfo
  {pages} {1} (\bibinfo {year} {1975})}\BibitemShut {NoStop}%
\bibitem [{\citenamefont {Musco}\ and\ \citenamefont
  {Miller}(2013)}]{Musco:2012au}%
  \BibitemOpen
  \bibfield  {author} {\bibinfo {author} {\bibfnamefont {I.}~\bibnamefont
  {Musco}}\ and\ \bibinfo {author} {\bibfnamefont {J.~C.}\ \bibnamefont
  {Miller}},\ }\href {\doibase 10.1088/0264-9381/30/14/145009} {\bibfield
  {journal} {\bibinfo  {journal} {Class. Quant. Grav.}\ }\textbf {\bibinfo
  {volume} {30}},\ \bibinfo {pages} {145009} (\bibinfo {year} {2013})},\
  \Eprint {http://arxiv.org/abs/1201.2379} {arXiv:1201.2379 [gr-qc]}
  \BibitemShut {NoStop}%
\bibitem [{\citenamefont {Harada}\ \emph {et~al.}(2016)\citenamefont {Harada},
  \citenamefont {Yoo}, \citenamefont {Kohri}, \citenamefont {Nakao},\ and\
  \citenamefont {Jhingan}}]{harada2016primordial}%
  \BibitemOpen
  \bibfield  {author} {\bibinfo {author} {\bibfnamefont {T.}~\bibnamefont
  {Harada}}, \bibinfo {author} {\bibfnamefont {C.-M.}\ \bibnamefont {Yoo}},
  \bibinfo {author} {\bibfnamefont {K.}~\bibnamefont {Kohri}}, \bibinfo
  {author} {\bibfnamefont {K.-i.}\ \bibnamefont {Nakao}}, \ and\ \bibinfo
  {author} {\bibfnamefont {S.}~\bibnamefont {Jhingan}},\ }\href@noop {}
  {\bibfield  {journal} {\bibinfo  {journal} {The Astrophysical Journal}\
  }\textbf {\bibinfo {volume} {833}},\ \bibinfo {pages} {61} (\bibinfo {year}
  {2016})}\BibitemShut {NoStop}%
\bibitem [{\citenamefont {Harada}\ \emph {et~al.}(2017)\citenamefont {Harada},
  \citenamefont {Yoo}, \citenamefont {Kohri},\ and\ \citenamefont
  {Nakao}}]{Harada:2017fjm}%
  \BibitemOpen
  \bibfield  {author} {\bibinfo {author} {\bibfnamefont {T.}~\bibnamefont
  {Harada}}, \bibinfo {author} {\bibfnamefont {C.-M.}\ \bibnamefont {Yoo}},
  \bibinfo {author} {\bibfnamefont {K.}~\bibnamefont {Kohri}}, \ and\ \bibinfo
  {author} {\bibfnamefont {K.-I.}\ \bibnamefont {Nakao}},\ }\href {\doibase
  10.1103/PhysRevD.96.083517} {\bibfield  {journal} {\bibinfo  {journal} {Phys.
  Rev. D}\ }\textbf {\bibinfo {volume} {96}},\ \bibinfo {pages} {083517}
  (\bibinfo {year} {2017})},\ \bibinfo {note} {[Erratum: Phys.Rev.D 99, 069904
  (2019)]},\ \Eprint {http://arxiv.org/abs/1707.03595} {arXiv:1707.03595
  [gr-qc]} \BibitemShut {NoStop}%
\bibitem [{\citenamefont {Carr}\ \emph {et~al.}(2021)\citenamefont {Carr},
  \citenamefont {Kohri}, \citenamefont {Sendouda},\ and\ \citenamefont
  {Yokoyama}}]{carr2020constraints}%
  \BibitemOpen
  \bibfield  {author} {\bibinfo {author} {\bibfnamefont {B.}~\bibnamefont
  {Carr}}, \bibinfo {author} {\bibfnamefont {K.}~\bibnamefont {Kohri}},
  \bibinfo {author} {\bibfnamefont {Y.}~\bibnamefont {Sendouda}}, \ and\
  \bibinfo {author} {\bibfnamefont {J.}~\bibnamefont {Yokoyama}},\ }\href
  {\doibase 10.1088/1361-6633/ac1e31} {\bibfield  {journal} {\bibinfo
  {journal} {Reports on Progress in Physics}\ }\textbf {\bibinfo {volume}
  {84}},\ \bibinfo {pages} {116902} (\bibinfo {year} {2021})}\BibitemShut
  {NoStop}%
\bibitem [{\citenamefont {Green}\ and\ \citenamefont
  {Malik}(2001)}]{Green:2000he}%
  \BibitemOpen
  \bibfield  {author} {\bibinfo {author} {\bibfnamefont {A.~M.}\ \bibnamefont
  {Green}}\ and\ \bibinfo {author} {\bibfnamefont {K.~A.}\ \bibnamefont
  {Malik}},\ }\href {\doibase 10.1103/PhysRevD.64.021301} {\bibfield  {journal}
  {\bibinfo  {journal} {Phys. Rev. D}\ }\textbf {\bibinfo {volume} {64}},\
  \bibinfo {pages} {021301} (\bibinfo {year} {2001})},\ \Eprint
  {http://arxiv.org/abs/hep-ph/0008113} {arXiv:hep-ph/0008113} \BibitemShut
  {NoStop}%
\bibitem [{\citenamefont {Hidalgo}\ \emph {et~al.}(2017)\citenamefont
  {Hidalgo}, \citenamefont {De~Santiago}, \citenamefont {German}, \citenamefont
  {Barbosa-Cendejas},\ and\ \citenamefont {Ruiz-Luna}}]{Hidalgo:2017dfp}%
  \BibitemOpen
  \bibfield  {author} {\bibinfo {author} {\bibfnamefont {J.~C.}\ \bibnamefont
  {Hidalgo}}, \bibinfo {author} {\bibfnamefont {J.}~\bibnamefont
  {De~Santiago}}, \bibinfo {author} {\bibfnamefont {G.}~\bibnamefont {German}},
  \bibinfo {author} {\bibfnamefont {N.}~\bibnamefont {Barbosa-Cendejas}}, \
  and\ \bibinfo {author} {\bibfnamefont {W.}~\bibnamefont {Ruiz-Luna}},\ }\href
  {\doibase 10.1103/PhysRevD.96.063504} {\bibfield  {journal} {\bibinfo
  {journal} {Phys. Rev. D}\ }\textbf {\bibinfo {volume} {96}},\ \bibinfo
  {pages} {063504} (\bibinfo {year} {2017})},\ \Eprint
  {http://arxiv.org/abs/1705.02308} {arXiv:1705.02308 [astro-ph.CO]}
  \BibitemShut {NoStop}%
\bibitem [{\citenamefont {Martin}\ \emph {et~al.}(2020)\citenamefont {Martin},
  \citenamefont {Papanikolaou},\ and\ \citenamefont {Vennin}}]{Martin:2019nuw}%
  \BibitemOpen
  \bibfield  {author} {\bibinfo {author} {\bibfnamefont {J.}~\bibnamefont
  {Martin}}, \bibinfo {author} {\bibfnamefont {T.}~\bibnamefont
  {Papanikolaou}}, \ and\ \bibinfo {author} {\bibfnamefont {V.}~\bibnamefont
  {Vennin}},\ }\href {\doibase 10.1088/1475-7516/2020/01/024} {\bibfield
  {journal} {\bibinfo  {journal} {JCAP}\ }\textbf {\bibinfo {volume} {01}},\
  \bibinfo {pages} {024} (\bibinfo {year} {2020})},\ \Eprint
  {http://arxiv.org/abs/1907.04236} {arXiv:1907.04236 [astro-ph.CO]}
  \BibitemShut {NoStop}%
\bibitem [{\citenamefont {Carrion}\ \emph
  {et~al.}(2021{\natexlab{a}})\citenamefont {Carrion}, \citenamefont {Hidalgo},
  \citenamefont {Montiel},\ and\ \citenamefont {Padilla}}]{reh5}%
  \BibitemOpen
  \bibfield  {author} {\bibinfo {author} {\bibfnamefont {K.}~\bibnamefont
  {Carrion}}, \bibinfo {author} {\bibfnamefont {J.~C.}\ \bibnamefont
  {Hidalgo}}, \bibinfo {author} {\bibfnamefont {A.}~\bibnamefont {Montiel}}, \
  and\ \bibinfo {author} {\bibfnamefont {L.~E.}\ \bibnamefont {Padilla}},\
  }\href {\doibase 10.1088/1475-7516/2021/07/001} {\bibfield  {journal}
  {\bibinfo  {journal} {JCAP}\ }\textbf {\bibinfo {volume} {07}},\ \bibinfo
  {pages} {001} (\bibinfo {year} {2021}{\natexlab{a}})},\ \Eprint
  {http://arxiv.org/abs/2101.02156} {arXiv:2101.02156 [astro-ph.CO]}
  \BibitemShut {NoStop}%
\bibitem [{\citenamefont {{de Jong}}\ \emph {et~al.}(2021)\citenamefont {{de
  Jong}}, \citenamefont {{Aurrekoetxea}},\ and\ \citenamefont
  {{Lim}}}]{2021arXiv210904896D}%
  \BibitemOpen
  \bibfield  {author} {\bibinfo {author} {\bibfnamefont {E.}~\bibnamefont {{de
  Jong}}}, \bibinfo {author} {\bibfnamefont {J.~C.}\ \bibnamefont
  {{Aurrekoetxea}}}, \ and\ \bibinfo {author} {\bibfnamefont {E.~A.}\
  \bibnamefont {{Lim}}},\ }\href@noop {} {\bibfield  {journal} {\bibinfo
  {journal} {arXiv e-prints}\ ,\ \bibinfo {eid} {arXiv:2109.04896}} (\bibinfo
  {year} {2021})},\ \Eprint {http://arxiv.org/abs/2109.04896} {arXiv:2109.04896
  [astro-ph.CO]} \BibitemShut {NoStop}%
\bibitem [{\citenamefont {Carrion}\ \emph
  {et~al.}(2021{\natexlab{b}})\citenamefont {Carrion}, \citenamefont {Hidalgo},
  \citenamefont {Montiel},\ and\ \citenamefont {Padilla}}]{Carrion:2021yeh}%
  \BibitemOpen
  \bibfield  {author} {\bibinfo {author} {\bibfnamefont {K.}~\bibnamefont
  {Carrion}}, \bibinfo {author} {\bibfnamefont {J.~C.}\ \bibnamefont
  {Hidalgo}}, \bibinfo {author} {\bibfnamefont {A.}~\bibnamefont {Montiel}}, \
  and\ \bibinfo {author} {\bibfnamefont {L.~E.}\ \bibnamefont {Padilla}},\
  }\href@noop {} {\  (\bibinfo {year} {2021}{\natexlab{b}})},\ \Eprint
  {http://arxiv.org/abs/2101.02156} {arXiv:2101.02156 [astro-ph.CO]}
  \BibitemShut {NoStop}%
\bibitem [{\citenamefont {Jedamzik}\ \emph {et~al.}(2010)\citenamefont
  {Jedamzik}, \citenamefont {Lemoine},\ and\ \citenamefont
  {Martin}}]{Jedamzik:2010hq}%
  \BibitemOpen
  \bibfield  {author} {\bibinfo {author} {\bibfnamefont {K.}~\bibnamefont
  {Jedamzik}}, \bibinfo {author} {\bibfnamefont {M.}~\bibnamefont {Lemoine}}, \
  and\ \bibinfo {author} {\bibfnamefont {J.}~\bibnamefont {Martin}},\ }\href
  {\doibase 10.1088/1475-7516/2010/04/021} {\bibfield  {journal} {\bibinfo
  {journal} {JCAP}\ }\textbf {\bibinfo {volume} {04}},\ \bibinfo {pages} {021}
  (\bibinfo {year} {2010})},\ \Eprint {http://arxiv.org/abs/1002.3278}
  {arXiv:1002.3278 [astro-ph.CO]} \BibitemShut {NoStop}%
\bibitem [{\citenamefont {Dalianis}\ and\ \citenamefont
  {Kouvaris}(2021)}]{Dalianis:2020gup}%
  \BibitemOpen
  \bibfield  {author} {\bibinfo {author} {\bibfnamefont {I.}~\bibnamefont
  {Dalianis}}\ and\ \bibinfo {author} {\bibfnamefont {C.}~\bibnamefont
  {Kouvaris}},\ }\href {\doibase 10.1088/1475-7516/2021/07/046} {\bibfield
  {journal} {\bibinfo  {journal} {JCAP}\ }\textbf {\bibinfo {volume} {07}},\
  \bibinfo {pages} {046} (\bibinfo {year} {2021})},\ \Eprint
  {http://arxiv.org/abs/2012.09255} {arXiv:2012.09255 [astro-ph.CO]}
  \BibitemShut {NoStop}%
\bibitem [{\citenamefont {Sasaki}\ \emph {et~al.}(2018)\citenamefont {Sasaki},
  \citenamefont {Suyama}, \citenamefont {Tanaka},\ and\ \citenamefont
  {Yokoyama}}]{Sasaki:2018dmp}%
  \BibitemOpen
  \bibfield  {author} {\bibinfo {author} {\bibfnamefont {M.}~\bibnamefont
  {Sasaki}}, \bibinfo {author} {\bibfnamefont {T.}~\bibnamefont {Suyama}},
  \bibinfo {author} {\bibfnamefont {T.}~\bibnamefont {Tanaka}}, \ and\ \bibinfo
  {author} {\bibfnamefont {S.}~\bibnamefont {Yokoyama}},\ }\href {\doibase
  10.1088/1361-6382/aaa7b4} {\bibfield  {journal} {\bibinfo  {journal} {Class.
  Quant. Grav.}\ }\textbf {\bibinfo {volume} {35}},\ \bibinfo {pages} {063001}
  (\bibinfo {year} {2018})},\ \Eprint {http://arxiv.org/abs/1801.05235}
  {arXiv:1801.05235 [astro-ph.CO]} \BibitemShut {NoStop}%
\end{thebibliography}%

\end{document}